%Paper: hep-th/9310151
%From: antoniad@orphee.polytechnique.fr
%Date: Fri, 22 Oct 1993 21:09:47 +0100

%         1         2         3         4         5         6         7
%1234567890123456789012345678901234567890123456789012345678901234567890123456789
% Equation labeling
\newskip\oneline \oneline=1em plus.3em minus.3em
\newskip\halfline \halfline=.5em plus .15em minus.15em
\newbox\sect
\newcount\eq
\newbox\lett
%newbox\notenum \setbox\notenum=\hbox{\footnotesize [00]\ \ }

\def\simlt{\mathrel{\lower2.5pt\vbox{\lineskip=0pt\baselineskip=0pt
           \hbox{$<$}\hbox{$\sim$}}}}
\def\simgt{\mathrel{\lower2.5pt\vbox{\lineskip=0pt\baselineskip=0pt
           \hbox{$>$}\hbox{$\sim$}}}}

\newdimen\short
\def\adv{\global\advance\eq by1}
\def\set#1#2{\setbox#1=\hbox{#2}}
\def\nextlet#1{\global\advance\eq by-1\setbox
                \lett=\hbox{\rlap#1\phantom{a}}}

\newcount\eqncount
%\newcount\sectcount
\eqncount=0
%\sectcount=0
%\def\sectadv{\global\advance\sectcount by1    }
%\def\secta{\global\advance\sectcount by1    }
%\def\equn{\global\advance\eqncount by1\eqno{(\copy\sect.\the\eqncount)} }
\def\equn{\global\advance\eqncount by1\eqno{(\the\eqncount)} }
\def\put#1{\global\edef#1{(\the\eqncount)}           }

\def\np{{\it Nucl. Phys.}}
\def\pl{{\it Phys. Lett.}}

\def\cKK{1}
\def\cGsw{2}
\def\cAbl{3}
\def\cAnt{4}
\def\cAqm{5}
\def\cKs{6}
\def\cDhvw{7}
\def\cGP{8}
\def\corb{9}
\def\cfterm{10}
\def\cPdg{11}
\def\cLep{12}
\def\cTas{13}

\magnification=1200
\hsize=6.0 truein
\vsize=8.5 truein
\baselineskip 14pt

\nopagenumbers

\rightline{CPTH-A257.0793}
\rightline{October 1993}
\vskip 1.0truecm
\centerline{\bf LIMITS ON EXTRA DIMENSIONS}
\centerline{\bf IN}
\centerline{\bf ORBIFOLD COMPACTIFICATIONS OF SUPERSTRINGS}
\vskip 1.0truecm
\centerline{{\bf I. Antoniadis} and {\bf K. Benakli}}
\vskip .5truecm
\centerline{{\it Centre de Physique Th{\'e}orique}
\footnote{$^*$}{\it Laboratoire Propre du CNRS UPR A.0014}}
\centerline{\it Ecole Polytechnique, 91128 Palaiseau, France}
\vskip 2.5truecm
\centerline{\bf ABSTRACT}
\vskip .5truecm

Perturbative breaking of supersymmetry in four-dimensional string
theories predict in general the existence of new large dimensions at the
TeV scale. Such dimensions can be consistent with perturbative
unification up to the Planck scale in a class of string models and open the
exciting possibility of lowering a part of the massive string spectrum at
energies accessible to future accelerators. The main signature is the
production of Kaluza-Klein excitations which have a very particular
structure, strongly correlated with the supersymmetry breaking mechanism.
We present a model independent analysis of the physics of these states in
the context of orbifold compactifications of the heterotic superstring. In
particular, we compute the limits on the size of large dimensions used to
break supersymmetry.

\hfill\break
%\vskip 3.5truecm
%\noindent CPTH-A257.0793
%\noindent October 1993
\vfill\eject

\footline={\hss\tenrm\folio\hss}\pageno=1

Our observable world is a four dimensional space-time and
there is no experimental evidence of the presence of extra dimensions.
However in trying to unify gravity with the other fundamental interactions,
we are lead to consider Kaluza-Klein (KK) type theories living in bigger
spaces [\cKK]. There, spacetime is assumed to be a product ${\cal M}_4
\times K$ of the Minkowski space ${\cal M}_4$ described by four non-compact
coordinates $x^{\mu}$ where ${\mu}=0,...,3$, with some internal space $K$
formed by new dimensions $X^i$ where $i=4,...,3+D$. In general, these
extra dimensions can be compactified to very small size, of the order of
the Plank length $M_p^{-1}$, so that they have no effect on present
experimental observations.

The most promising of such extensions of general relativity is superstring
theory which provides the only known example of consistent quantum gravity
[\cGsw]. One of the main problems of this theory is to break spacetime
supersymmetry at energies close to the electroweak scale in order to
protect the gauge hierarchy. In contrast to the situation in field theory
where one has the freedom to introduce arbitrary soft breaking masses, in
string theory the scale of supersymmetry breaking is not a new independent
parameter. In perturbation theory, it has to be of the same order with the
inverse size of some internal dimension(s) [\cAbl]. This leads to a new
interest in models with at least one large extra dimension lying in an
energy domain where experiments can be performed.

In quantum field theories such models have serious problems: they are not
renormalizable, and moreover the coupling constants grow very rapidly to
non-perturbative domain above the compactification scale. In the
framework of string theory, these problems can be avoided in a class of
four-dimensional models which include orbifold compactifications [\cAnt].
The crucial property is that all Kaluza-Klein modes which are
associated with the large internal coordinate are organized in multiplets
of $N=4$ (spontaneously broken) supersymmetry, leading to cancellations of
large radiative corrections among particles of different spins. In
particular the evolution of gauge couplings remains logarithmic, as in a
four-dimensional theory, up to the Planck scale.

An explicit realization of the minimal supersymmetric standard model in
the context of this perturbative breaking mechanism was studied recently
and it was found to be extremely restrictive [\cAqm]. Quarks and leptons,
unlike gauge bosons, should be identified with ``twisted" states having no
KK-excitations. On the other hand there is a simple choice for the
Higgs-sector where the second Higgs doublet can be identified with the first
KK-excitation of the first doublet having opposite chirality. In this way,
a Higgs-mixing term appears in the effective superpotential with a
coefficient $\mu$ also related to the compactification scale. Another
important feature of this mechanism is the absence of quadratic
divergences in the vacuum energy. This allows the possibility of
generating the large hierarchy between the compactifiaction and Planck
lengths radiatively, by minimizing the full one loop effective potential.

It is then important to extract experimental limits on the size of
new large dimensions. In this work we obtain upper bounds in the
context of orbifold compactifications where exact calculations can be
performed. In particular we restrict to the class of models studied in
[\cAqm] and based on ${\bf Z}_N$ ($N>2$) orbifolds which require at least
two large internal coordinates. The main signature of the existence of
these dimensions is the appearance of KK-excitations of gauge bosons and
higgses (as well as their superpartners). All these massive states are
instable and can desintergrate into massless quarks and leptons (twisted
states). This is due to the absence of any conserved quantum numbers
associated to the massive modes, in contrast to ordinary Kaluza-Klein
theories. This property seems to be related to the chiral character of the
compactification and it implies that the theory remains four dimensional at
all energy scales below $M_p$, as the massive states cannot propagate
longer than their short lifetime. However, their exchange gives rise at
low energy to effective four-fermion interactions  which modify known cross
sections or lead to new physical processes. We compute the strength of such
interactions and use the present experimental limits to get bounds on the
size of extra dimensions.

Similar limits have been derived in the past by a model
independent analysis of various experimental data [\cKs]. In the
general case there are many parameters related to the couplings of massive
to massless states, and thus the simplification of taking into account only
the first KK-excitation was made. As we will see, this is not always
valid. In fact assuming equal couplings for all massive modes, this
approximation is legitimate only in the case of one extra dimension, since
the summation over the infinite tower of KK-states leads to a small
correction [\cAqm]. However, in the case of more than one internal
coordinates the sum diverges and the approximation breaks down. To get a
finite answer the couplings should decrease with the mass and the result in
general depends on their exact form.

For pedagogical reasons, we first recall these computations in the
simple case where the large internal dimension is a circle of radius $R$.
Then any field $\Phi$ satisfies the periodicity condition
$$
\Phi (x,X^4+ 2{\pi}R)= \Phi (x,X^4) \ ,
\equn\put\period
$$
with $X^4$ the circle coordinate, and it has the Fourier expansion
$$
\Phi (x,X^4)= {\sum}_n \Phi_n (x)  e^{i{ n X^4 \over R}}\ .
\equn\put\fourier
$$
This means that the momentum $P_4$ associated to $X^4$ is quantized in
integer multiples of $1\over R$. Therefore the particle spectrum forms a
tower of KK-states with the same quantum numbers except their masses
which are given by:
$$
{{m_n}^2 = {m_0}^2 + {{n^2} \over {R^2}}}\ ,
\equn\put\kkmass
$$
where $m_0$ is the higher-dimensional mass. For compactifications where
$P_4$ is not conserved, the exchange of massive gauge bosons
beetween quarks and leptons gives rise at low energy to an effective
interaction described by dimension six four-fermion operators of the
form:
$$\eqalign{
L_n &= -{g_n^2 \over 2m_n^2}{\cal O}_{\psi\psi\psi\psi}\cr
{\cal O}_{\psi\psi\psi\psi} &=[{\eta_{LL}}({{\bar
\psi}_L{\gamma^\mu}{\psi}_L})^2 +
 {\eta_{RR}}({{\bar \psi}_R{\gamma^\mu}{\psi}_R})^2 +
2{\eta_{RL}}({{\bar \psi}_R{\gamma^\mu}{\psi}_R})
            ({{\bar \psi}_L{\gamma^\mu}{\psi}_L})]\ ,\cr}
\equn\put\leff
$$
where ${\eta_{IJ}}$ parametrize the relative strength of
left-left $(LL)$, right-right $(RR)$ and left-right $(LR)$ interactions,
while $\eta_{IJ}g_n^2$ are the corresponding coupling constants. In
{\leff} we used for simplicitly the same generic symbol $\psi$ for all
fermions which in principle can be different species. Note that the
exchange of massive scalars with Yukawa couplings lead also to effective
four-fermion interactions which can be put in the same form as {\leff} by
appropriate Fierz-transformations.

Assuming $g_n=g$ independent of $n$ and using the mass formula {\kkmass},
one can perform the sum over all massive modes with the same quantum
numbers ($|n| \ge 1$) to get:
$$
L_{eff}=-{g^2 \over 2}({\pi \over  R m_0 } \coth{(\pi m_0 R )}-{1 \over
R^2 m_0^2}) R^2 {\cal O}_{\psi\psi\psi\psi}\ .
\equn\put\sleff
$$
This result remains valid for large $R$ even if the condition of equal
couplings is recovered only asymptotically as $R\to\infty$, provided
$g_n/{\sqrt{|n|}}$ decreases with $|n|$ fast enough that the series
converges. In the case of two such large extra dimensions, the same
considerations lead to summing on two integer numbers associated to the
two quantized momenta but this sum diverges logarithmically. A non-trivial
dependence of the couplings on $R$ is then expected to arise from a
meaningfull theory in order to regularize the divergent expression.
\vskip 1.0cm
\centerline{\bf PHYSICS OF KALUZA-KLEIN STATES IN ORBIFOLD MODELS}
\vskip 0.5cm

We turn now to the case of orbifold compactifications of the heterotic
string theory [\cDhvw]. In general, one starts with four-dimensional
vacuua having $N=4$ spacetime supersymmetry, like toroidal
compactifications, and divide by a discrete symmetry group of the internal
space to obtain a chiral spectrum with $N=1$ supersymmetry. The latter
implies a complexification of the internal coordinates, while to realize a
${\bf Z}_N$ symmetry with $N>2$ one needs at least one complex coordinate
$X$ compactified on a two-dimensional lattice defined by:
$$
X \equiv X+ 2\pi R (n_1+ n_2\theta ) \quad ; \quad
\theta = e^{{2\pi}i\over N}\ .
\equn\put\lat
$$
Note that in the limit $R\to\infty$ both dimensions become large. The
associated Hilbert space contains two sectors of states:

\noindent
-The untwisted sector where, going around the string, $X$ is periodic
up to a lattice translation. This sector contains the states of the
toroidal compactification which are also invariant under the orbifold
projection ${\bf Z}_N$. Their masses are given in {\kkmass} with $n$
replaced by a two-component vector ${\vec n}=(n_1,n_2)$ or, equivalently,
by a complex number $n \equiv n_1 + n_2 \theta$. Here, we neglect the
string winding modes, whose masses are quantized in units of $R$, and
thus they become irrelevant in the limit of large $R$. In the class of
models we consider, the KK-excitations are organized in multiplets of
$N=4$ supersymmetry. Moreover, all states of the toroidal
compactification are in representations of the ${\bf Z}_N$ discrete
group. These representations contain states with different quantum
numbers, since ${\bf Z}_N$ transformations act also on the 2d fermionic
superpartner of the internal coordinate $X$. As a result, on the
``massless" spectrum (${\vec n}=0$), the orbifold acts as a chiral
projection which also breaks the gauge group. However the massive
KK-states (${\vec n}\ne 0$) have an additional ${\bf Z}_N$-degeneracy
associated to the lattice, and all quantum numbers of the $N=4$ theory
remain present even after the orbifold projection.

\noindent
-The twisted sectors $\{\theta^k\}$ where, going around the string, the
internal coordinate $X$ picks up a phase $\theta^k$ (up to a lattice
translation). The string center of mass is sitting on a fixed point of
the lattice under the transformation $\theta^k$ and the corresponding
states do not carry internal momenta. The presence of these states, which
do not have KK-excitations, is required from the consistency of the string
theory and is related to its chiral character.

The perturbative mechanism of spontaneous supersymmetry breaking in these
models, at a scale proportional to $1/R$, leads to a simple pattern of
soft breaking masses at the tree-level [\cAqm]. Only the fermions from the
untwisted sector aquire a common mass-shift while untwisted bosons and
twisted matter remain intact. In particular there is a universal gaugino
mass, since gauge bosons arise in the untwisted sector. Moreover, quarks
and leptons must be identified with twisted states having no
KK-excitations. On the other hand, the higgses can be chosen either in
the untwisted or in the twisted sector. The former choice also
offers an interesting explanation for the origin of the second Higgs
doublet, present in any supersymmetric extension of the standard model, as
well as its mixing with the first doublet, which forbids the presence of an
unwanted electroweak axion. In fact starting with one massless Higgs
doublet at the level of the supersymmetric theory, one can identify the
second doublet with its lowest KK-excitation of opposite chirality (and
opposite hypercharge) [\cAqm]. For the purpose of this work, the effects
of supersymmetry breaking are not important and will be neglected in the
rest of our analysis.

In the following, we study the main characteristics of the physics of
KK-excitations. As already mentioned above, these exhibit a larger
symmetry which is present even before the orbifold projection, at the level
of the $N=4$ supersymmetric theory. An $N=4$ multiplet contains one vector
boson, four two-component fermions and six real scalars, or equivalently,
one vector and three chiral $N=1$ multiplets. The interactions between the
massive and massless untwisted states are therefore determined entirely
by the $N=4$ symmetry once the gauge group is specified. Considering
only the excitations associated to the large complex dimension $X$, and
neglecting gravitational interactions, this infinite dimensional
gauge group of the $N=4$ theory, in the large radius limit, is given by
[\cGP]:
$$
[T^a_n , T^b_m] = f^{ab}{}_c T^c_{n+m}\ ,
\equn\put\exsym
$$
where $T^a_0$ are the generators of the gauge group $G$ of the
massless states and $f^{ab}{}_c$ are the corresponding structure
constants. $T^a_n$  with $n\ne 0$ are associated to the
massive modes and they extend the gauge algebra to an infinite dimensional
Kac-Moody type symmetry.

The discrete orbifold group ${\bf Z}_N$ corresponds to an automorphism of
the algebra {\exsym}. In an appropriate basis where ${\bf Z}_N$
transformations are diagonalized, the generators $T^a_0$ are divided
into $N$ sets of eigenvectors with eigenvalues $\theta^k$
($k=0,1,...,N-1$). The orbifold projection breaks the gauge symmetry, since
it keeps in the massless spectrum only the invariant set of generators
which form a maximal subgroup $H$ of $G$. At the massive levels, the
orbifold group acts also on the internal momentum taking $n$ to
$\theta^k n$. This allows to construct ${\bf Z}_N$ invariant
excitations for all the generators of $G$, which are given by the following
linear combinations:
$$
{\tilde T}^k_n\equiv {1\over N}\sum_{r=0}^{N-1}
\theta^{kr} T^k_{\theta^r n}\ ,
\equn\put\gen
$$
where $T^k_0$ denote the generators of $G$ transforming with a
phase $\theta^k$ under ${\bf Z}_N$, and the corresponding gauge indices
$\{ a\}$ were omitted for notational simplicity. ${\tilde T}^k_n$
satisfy the algebra:
$$
[{\tilde T}^k_n , {\tilde T}^l_m] =
{1\over N}\sum_{r=0}^{N-1}\theta^{lr}
{\tilde T}^{k+l}_{n+\theta^{r}m}\ ,
\equn\put\sym
$$
where the structure constants of $G$ were also omitted in the r.h.s.

The exchange of the massive states associated to the generators of $G/H$,
${\tilde T}^k_n$ with $k\ne 0$, could give rise to new low energy
interactions which are in general subject to more stringent experimental
constraints. In particular they could lead to phenomenological problems
like fast proton decay or flavor changing neutral currents. A part of
such potential problems are already avoided since we identified quarks and
leptons with twisted states which have no KK-excitations.

In the minimal case where the unbroken low energy group $H$ is just the
Standard model $SU(3)_c\times SU(2)_w\times U(1)_Y$ with the Higgs
doublets in the untwisted sector, there are only three distinct
possibilities for the gauge group $G$ of the $N=4$ supersymmetric theory,
before the orbifold projection:
$$
G=SU(3)_c\times SU(3)\quad ,\quad
G_2\times SU(3)\quad ,\quad F_4\ ,
\equn\put\groups
$$
where $F_4$ has two different embeddings. The other rank-four groups which
have $SU(3)\times SU(2)\times U(1)$ as maximal subgroup were excluded
because they do not contain in their adjoint representations a Higgs like
doublet. Moreover $SU(3)_c\times G_2$ is not considered because it leads to
color singlets KK-modes with fractional electric charges.
$SU(3)_c\times SU(3)$ is the minimal choice which contains only the
excitations of one Higgs doublet in addition to those of the Standard model
gauge bosons. In fact the other two groups in {\groups} are simple
extensions of this case since they contain $[SU(3)]^2$ as subgroup.

Note that $G$ is not a real grand-unified group above the
decompactification scale because the matter generations are not present at
the level of the $N=4$ theory. They appear as twisted states after the
orbifold projection and they cannot in general be embedded in
representations of $G$. Consider for instance the simple ${\bf Z}_3$
orbifold which breaks $G=E_8$ down to $H=E_6\times SU(3)$. The matter
multiplets in the untwisted sector transform as $(27,3)$ which is part of
the adjoint of $E_8$, while the twisted states transform as $(27,1)$ or
$(1,3)$ which cannot be embedded in any representation of $E_8$.

The presence of twisted states introduces new interactions which are not
constrained by the $N=4$ symmetry. However, it is easy to show by group
theory arguments that ordinary matter fermions couple only to the vector
excitations of the Standard model gauge bosons through gauge interactions,
and to the scalar excitations of the Higgs doublet through Yukawa
interactions. Thus, there are no new effective four-fermion operators and
it is sufficient to examine the modifications to known low energy
cross sections. Moreover, the Higgs excitations are not expected to
provide important model-independent limits as long as the lowest Higgs
scalar has not yet been observed. Therefore, our bounds on the size $R$
will be obtained from the exchange of the excitations of gauge bosons.

The resulting four-fermion interaction is of the form {\leff} where the
index $n$ is replaced by a two-dimensional vector ${\vec n}$ associated to
the ${\vec n}$-th KK-excitation of one of the Standard model gauge bosons.
Furthermore, the sum must be done over all ${\bf Z}_N$ invariant linear
combinations $\{ {\vec n}\}$ corresponding to the generators ${\tilde
T}^0_n$ defined in {\gen}. The coupling constant $g_{\{ {\vec n}\}
}$ of the $\{ {\vec n}\}$-th KK-mode (properly normalized) to two massless
states from the twisted sectors $\{\theta^k,\theta^{-k}\}$ is [\corb]:
$$
g_{\{ {\vec n}\} }=g e^{i\pi{\vec\epsilon}\cdot{\vec n}}
\delta^{-{{\vec n}^2\over 2R^2}}\ ,
\equn\put\gn
$$
where our mass units are defined by setting the string tension
$\alpha'=2$, and $\delta$ is given by $\ln\delta =
2\psi(1)-\psi(k/N)-\psi(1-k/N)$ with specific values $\delta = $ 16, 27,
64, 432 for $k/N = $ 1/2, 1/3, 1/4 and 1/6, respectively. The phase factor
in {\gn} depends on the fixed point associated to the twisted state under
consideration; its position is given by ${\vec\epsilon}\pi R$. The
strongest bound comes when all four fermions arise at the same fixed point,
in which case the phases disappear in the amplitude. Then, the sum over the
massive ${\bf Z}_N$ invariant states, in the region where ${m_0 R} < 1$,
gives:
$$
L_{eff}=-
{g^2 \over 2}(c_1 \ln{( R^2 )}+ c_2 ) R^2 {\cal O}_{\psi\psi\psi\psi}\ ,
\equn\put\nsleff
$$
where $c_1={2 \pi \over {3 \sqrt{3}} }$, ${ \pi \over 4 }$, ${ \pi \over{3
\sqrt{3}}}$ for ${\bf Z}_3$, ${\bf Z}_4$ and ${\bf Z}_6$, respectively. The
constant $c_2$ has a value of order unity and it is negligible, in the
large radius limit, compared to the $\ln{(R^2)}$ term. The same computation
for the case of a ${\bf Z}_2$ twist with just one large dimension leads to:
$$
L_{eff}=-
{g^2 \over 2}{\pi^2 \over 6} R^2 {\cal O}_{\psi\psi\psi\psi}\ .
\equn\put\unleff
$$
Note that in the case where the two pairs of fermions correspond to
different fixed points one can take the limit $R\to\infty$ in the
expression of the couplings {\gn}, since the sum now converges because of
the presence of the phase factor proportional to the relative distance
between the two fixed points. The strength of the four-fermion operator
behaves now as $R^2$ like in the one-dimensional case.

In {\nsleff-\unleff}, $g$ is the tree level gauge coupling constant at
the string scale. However in order to compare with experiments,
one has to compute the renormalized four-fermion interaction at low
energies. The massive vector bosons exchanged in these processes are
associated to spontaneously broken gauge symmetries via a stringy Higgs
mechanism [\cGP]. Because of the corresponding Ward identities, the
radiative corrections to the propagators of the external fermions cancel
against the corrections to the interaction vertices (for an appropriate
choice of gauge). Thus, the only contribution to the renormalization of
$L_{eff}$ comes from the radiative corrections to the propagators of the
exchanged gauge bosons. These propagators are the sum of two terms, one
proportional to the four dimensional flat metric $g_{\mu \nu}$ and
another proportional to the corresponding momenta $p_{\mu} p_{\nu}$. The
latter does not contribute to the amplitude because of the conservation of
the external currents.

Let us then denote by $g_{\mu \nu}S(p,R,\delta)$
the sum of all tree level vector boson propagators weighted by a factor
$\delta^{-{{\vec n}^2\over R^2}}$ from the interaction vertices {\gn}.
Neglecting $m_0$, $S(p,R,\delta)$ is the sum of the massless gauge boson
propagator $1\over p^2$ with a correction $\delta S$ appearing in the
effective four-fermion interaction $L_{eff}$:
$$
S(p,R,\delta)= {1 \over p^2} + \delta S\ ,
\equn\put\Sbar
$$
where
$$
\delta S= -({\pi^2 \over 6}+ \cdots ) R^2
\equn\put\Sbara
$$
in the one dimensional example, while in the two dimensional case
described above,
$$
\delta S=- (c_1 \ln{( R^2 )} + c_2 + \cdots ) R^2 \ ,
\equn\put\Sbarb
$$
where the dots stand for terms which are vanishing in the limit
$pR\rightarrow 0$.

The one loop self-energy correction has the form $g^{\mu\nu}I_b (p^2)
= g^{\mu \nu}p^2 g^2 b L(p^2)$, where $L(p^2)\equiv \ln{p^2\over M_p^2}$
and $b$ is the $\beta$-function coefficient dependent on the massless modes
that propagate in the loop. The massive states do not contribute because of
the $N=4$ supersymmetry. When supersymmetry is broken, one expects small
modifications due to threshold corrections of massive particles. The
renormalized one loop effective propagator is obtained by summing over all
one loop bubles. In the simplest case where all twisted states come from
the same fixed point, the result is:
$$
S_{ren}(p,R,\delta)= {S(p,R,\delta) \over {1- {1\over p^2}I_b(p^2) -
\delta S(p,R,\delta) I_{b_T}(p^2)}}\ ,
\equn\put\Sren
$$
where $b_T$ denotes the contribution to the one loop $\beta$-function
of the twisted states only. Using {\Sbar}, the above expression takes
the form:
$$
g^2 S_{ren}(p,R,\delta)=
{g^2(p^2)  \over p^2} +{g^4(p^2) \over g^2_U(p^2)} {\delta S  \over
{1-p^2 \delta S({g^2(p^2) \over g^2_U(p^2)}-1)}}\ ,
\equn\put\Srenf
$$
where $g^2(p^2) ={ g^2 \over {1- g^2 b L(p^2) }}$ is the running
coupling constant associated to the massless gauge boson, and $g^2_U(p^2)$
is the effective coupling which takes into account only the contribution of
the massless untwisted states\footnote{$^1$}{Note that $g^2_U(p^2)$ should
not be considered as some physical effective coupling; it can even take
negative values for non abelian groups.}. The first term in {\Srenf} can be
identified as the contribution to the amplitude of the massless gauge boson
with the usual running coupling constant, while the second term contains
the contribution of the tower of KK-modes.

Before obtaining the numerical bounds, in order to justify the effective
field theory approach used above, we present a direct derivation of the
effective four-fermion interactions in string theory.
\vskip 1.0cm
\centerline{\bf STRING COMPUTATION OF FOUR-FERMION OPERATORS}
\vskip 0.5cm

In string theory the emission or absorbtion of a fermionic state is
described by the appropriate Ramond vertex operator $V^R(z,\bar z)$
where $z$ and $\bar z$ are the world-sheet coordinates which must be
integrated on. In the heterotic string, this operator takes the form (in
the -1/2 ghost picture) [\corb]:
$$
V^R_{\pm}(z,\bar z) = e^{-{\varphi}/2}(z) S_{\pm}(z) \Psi_{\pm}(z,\bar z)
e^{i p_{\mu} x^{\mu}}\ ,
\equn\put\Vert
$$
where $\varphi$ is the bosonized super-reparametrization ghost, $S_{\pm}$
is a four-dimensional spin field with ${\pm}$ the chirality index, and
$\Psi_{\pm}(z,{\bar z})$ is a Ramond field of the internal
(super)-conformal field theory. After bosonizing the four fermionic
coordinates in terms of two free 2d scalars $\phi_{1,2}$, one has:
$$
S_+(z) = e^{\pm{i\over 2}(\phi_1 +\phi_2)}\ ,\quad
S_-(z) = e^{\pm{i\over 2}(\phi_1 -\phi_2)}\ .
\equn\put\defS
$$
Moreover, in the case of orbifolds when the fermions are in the twisted
sector, the internal part of the vertex, $\Psi_{\pm}$, becomes:
$$
\Psi_{\pm}(z,\bar z)= {\prod_{j=1}^3}
{\sigma^{(j)}_{{k_j \over N},\epsilon_j} (z,\bar z)}
e^{\pm i ({k_j \over N}-{1 \over 2}) H_j(z)} {\bar s}(\bar z) \,
\equn\put\dPsi
$$
where $j$ labels the three internal planes twisted by $e^{\pm 2i\pi
k_j/N}$ ($k_j=1,...,N$), and $\sigma^{(j)}_{{k_j \over N},\epsilon_j}$ are
the corresponding twist operators associated with the fixed points
$\epsilon_j$. $H_j$ are three free scalars which bosonize the three
complex internal fermionic coordinates, while $\bar s$ is an additional
right-moving part of the vertex operator containing the gauge group
dependence. Finally, space-time supersymmetry requires $\sum_j k_j=N$.

To retrieve for instance an effective four-fermion interaction of the form
\break $({\bar\psi}_0\gamma^\mu \psi_1) ({\bar\psi}_2\gamma_\mu \psi_3 )$,
we need to compute the following amplitude:
$$
{\cal A}=\int d^2z <{V^R_{0+}(0)  V^R_{1-}(z,\bar z) V^R_{2+}(1)
V^R_{3-}(\infty)}>\ ,
\equn\put\Vcorx
$$
where $V^R_i$ are the vertices corresponding to $\psi_i$. In {\Vcorx}, as
usual, we have used the global conformal invariance $SL(2,{\bf C})$ of the
world-sheet sphere to fixe the positions of three vertices at $0$, $1$, and
$\infty$. We will derive the effective four-fermion operator generated by
the exchange of massive string states in the $s$-channel. Consequently, we
can restrict the $z$-integration inside  a disk of radius one ($|z| \le
1$), since the integration outside the disk corresponds by duality to the
$u$-channel exchanges.

The non-trivial part of the amplitude {\Vcorx}, involves the correlator of
four twist operators [\corb]:
$$
<{\sigma^{(j)}_{{k_j \over N},0} (0)}
{\sigma^{(j)}_{1-{k_j \over N},\epsilon_1} (z, \bar z)}
{\sigma^{(j)}_{{l_j \over N},{\epsilon_1+\epsilon_3}} (1)}
{\sigma^{(j)}_{1-{l_j \over N},\epsilon_3} (\infty)}>\ ,
\equn\put\Ztw
$$
where we have chosen two pairs of opposite twists, since we need to
describe the exchange of massive gauge bosons which come in the untwisted
sector. In {\Ztw} we have also chosen the first fixed point to sit at the
origin. To illustrate the ideas we present the computation in the simplest
case of a ${\bf Z}_2$ twist, while the generalization to ${\bf Z}_N$ is
straightforward. Then, there are two fixed points for each dimension
parametrized by a number $\epsilon =0,1$ corresponding to the position
$\epsilon\pi R$. Furthermore all twists are identical,
${k_j/N}={l_j/N}=1/2$. To compute the correlation function {\Ztw}, one goes
from the sphere where the twisted quantities are multivalued to a torus
where the functions are well defined [\corb]. This torus is described by a
complex parameter $\tau = \tau_1 + i\tau_2$ ($\tau_2 \ge 0$) related to
the sphere coordinate $z$ by:
$$
z=\Bigl({\theta_2 (\tau )\over \theta_3 (\tau )} \Bigr)^4 \ ,
\equn\put\nVert
$$
where $\theta_i$ are the usual Jaccobi-Theta functions.

Using the form of
the fermion vertices {\Vert-\dPsi}, a straightforward computation of the
amplitude {\Vcorx} gives:
$$
\eqalign {{\cal A}=g^2 \int d^2z  &|z|^{-2 + s}  |1-z|^{-2 + t}
\{ (1-z)({\bar\psi}_0\gamma^\mu\psi_1) ({\bar\psi}_2\gamma_\mu\psi_3)
- z({\bar\psi}_0\gamma^\mu\psi_3) ({\bar\psi}_2\gamma_\mu\psi_1)\} \cr
&\times {1 \over | \theta_3 (\tau )|^{2d}} Z_R(\tau, R) Z_{int}(\tau )\
,\cr} \equn\put\namp
$$
where $Z_R$ contains the classical contibution of the lattice with the
large dimension(s) and $Z_{int}$ stands for the remaining internal part
(contribution of the other twisted coordinates).
$$
Z_R = {R^d \over \tau_2^{d/2}} \sum_{u,v}
e^{- {|u \tau + v|^2 \over {4 \pi \tau_2}}}
\equn\put\Zdef
$$
where $u=\pi R (2{\vec m}+{\vec \epsilon}_1)$, $v=\pi R (2{\vec
n}+{\vec \epsilon}_3)$, and the sum is over the integer components of the
$d$-dimensional vectors ${\vec m},{\vec n}$; $d=1,2$ in the case of one or
two large dimensions, respectively.

In the limit $R \to \infty$ there are three possible contributions to the
sum {\Zdef} which are not exponentially suppressed:
$$
\{u=0 \quad , \quad {\tau_2 \to \infty}\} \quad , \quad
\{v=0 \quad , \quad {\tau \to 0}\} \quad {\rm and} \quad \{ u=v=0\} \ .
\equn\put\ucond
$$
The condition $u=0$ implies ${\vec \epsilon}_1= 0$, while the condition
$v=0$ implies ${\vec \epsilon}_3= 0$. This corresponds to the case where
the four fermion vertices contain two pairs of twist-antitwist operators
situated at two fixed points, $0$ and $\epsilon_3 \pi R$ ($u=0$) or
$\epsilon_1 \pi R$ ($v=0$).

The limit ${\tau_2 \to \infty}$ is equivalent to ${z \to 0}$ where the
amplitude is factorized into the exchange of KK-states in the $s$-channel.
In fact, in this limit $\theta_3\to 1$, $Z_{int}$ goes to a constant,
while $\pi\tau_2 \sim -\ln {{|z| \over \delta}}$ with
$\delta =16$. After a Poisson ressumation in ${\vec n}$, one finds:
$$
Z_R \sim \sum_{\vec n} (-)^{{\vec n}\cdot {\vec \epsilon}_3} e^{-\pi
\tau_2 {{\vec n}^2 \over R^2}} \sim \sum_{\vec n} (-)^{{\vec n}\cdot {\vec
\epsilon}_3} ({|z| \over \delta})^{{\vec n}^2 \over R^2}\ .
\equn\put\Zres
$$
The limit ${\tau_2 \to 0}$ implies $z\to 1$ which corresponds to the
$t$-channel factorization. It can be obtained from the previous case by the
transformation $\tau\to -1/\tau$ sending $z\to (1-z)$ and $s\to t$. The
last case of {\ucond} can contribute only when $\epsilon_1=\epsilon_3=0$
implying that all four twist-operators come from the same fixed point.

Putting together all three contributions, one finds that in the large
radius limit the amplitude {\Vcorx} becomes:
$$
\eqalign{ {\cal A} &\sim g^2 ({\bar\psi}_0\gamma^\mu\psi_1)
({\bar\psi}_2\gamma_\mu\psi_3) \delta_{\epsilon_1 ,0} \int d^2z
|z|^{-2 + s} \{ \sum_{\vec n} (-)^{{\vec n}\cdot {\vec \epsilon}_3}
({|z| \over \delta})^{{\vec n}^2 \over R^2}\cr &\qquad \qquad\qquad\qquad
\qquad\qquad + \delta_{\epsilon_3 ,0} {R^d \over \tau_2^{d/2}}
( |1-z|^{-2+t} (1-z) {1\over |\theta_3 (\tau )|^{2d}} Z_{int}(\tau ) - 1)
\} \cr &- (s\leftrightarrow t , 1\leftrightarrow 3)\ , \cr }
\equn\put\nampas
$$
where we normalized $Z_{int}(i\infty )=1$. After integration, the first
term in {\nampas} gives at low energy:
$$
g^2 ({\bar\psi}_0\gamma^\mu\psi_1) ({\bar\psi}_2\gamma_\mu\psi_3)
({1\over s} + R^2 \sum_{\{ {\vec n}\}}(-)^{{\vec n}\cdot {\vec \epsilon}_3}
{{\delta}^{-{{\vec n}^2 \over R^2}} \over {\vec {n}}^2})\ ,
\equn\put\Leffa
$$
which reproduces the field theory result. The same expression can also be
obtained for general ${\bf Z}_N$ twists. The $1/s$ pole in {\Leffa}
corresponds to the exchange of massless gauge bosons, while the sum
represents the contribution of the massive KK-modes which was
computed in {\nsleff} and {\unleff} (for ${\vec \epsilon}_3 =0$ and the
corresponding values for $\delta$). The second term in {\nampas} which
behaves like $R^d$ takes into account the contribution of KK-excitations
of the massive string states. However its contribution is subleading in all
cases, and thus, it can be neglected.

In our previous field-theoretical analysis we studied the low energy
effects of the ``light" KK-modes which are associated to the massless
states. In the class of models we examined above (see discussion after eq.
{\groups}), these states do not lead to any new important interactions, but
their main effect consists in modifying known cross sections which we
analyzed by computing the strength of the effective four fermion operators.
However in any string model there are also superheavy states which may
generate new interactions whose couplings are normally suppressed by
powers of the Planck mass. These states have also a tower of KK-excitations
which may introduce large corrections invalidating the power suppression,
when a compactification radius becomes large. For instance, if $Z_{int}$
vanishes at all boundaries of $z$ (0, 1 and $\infty$), there is only one
possible contribution to the amplitude {\namp}, arising from the third
case in {\ucond} which leads to the second term of ``stringy" origin in the
integrand {\nampas}. Note however that in contrast to supersymmetric
D-terms we examine here, non-renormalizable F-terms are exponentially
suppressed in the large radius limit [\cfterm].

\vskip 1.0cm
\centerline{\bf NUMERICAL RESULTS}
\vskip 0.5cm

We now derive the numerical bounds. As explained above, these bounds can
be mainly obtained from modifications of low energy cross sections due to
the effective four-fermion interactions of the form {\unleff}, {\nsleff}
induced by the exchange of the KK-excitations of gauge bosons. The
presence of similar effective operators have been investigated in searches
for quark and lepton compositeness [\cPdg]. The best limits come from
$e^+e^-\rightarrow l^+l^-$ experiments with $l=e,\mu, \tau$. In our case,
the strength of these effective operators depends on the relative positions
of the fixed points associated to the external fermions. In fact, the
strongest bound is obtained when the four fermions are identified with
twisted states sitting at the same fixed point, giving rise to the
additional factor $\ln (R^2)$ in the case of two large dimensions. It turns
out that in the class of models we discussed previously, the assumption
that the Higgs field comes from the untwisted sector implies that the
Yukawa couplings  are exponentially suppressed in $R^2$ when the two
twisted fermions are in two different fixed points. Thus, to obtain
sensible quark and lepton masses, the left $SU(2)$-doublet and the
corresponding right singlets must always come from the same fixed point.
As a result, in our case, the strongest model independent bound is obtained
from $e^+e^-\rightarrow e^+e^-$ processes.

Here, we use the data of TASSO collaboration at PETRA which provides
actually the strongest limits on compositeness scale
[\cPdg]\footnote{$^2$}{LEP data, recently analyzed, provide bounds
for compositeness scale of the same order of magnitude [\cLep].}. Their
numerical values of differential cross-sections, given in ref.[\cTas],
allowed us to performe a $\chi^2$ fit to the renormalized amplitude
{\Srenf}. In our fit, we used the values $\alpha_{em}(M_Z)= 1/127$,
$\sin^2\theta_W(M_Z)=0.233$, and putted back our mass units from the
relation $\alpha'={4\over\alpha}M_p^{-2}$ where $\alpha=g^2/4\pi$ is the
string coupling constant. The determination of the value of $\alpha$, which
is also necessary for the computation of the effective coupling $g_U^2$, is
related to the problem of the unification scale. For the numerical values
given below, we used  $\alpha\sim 0.4$ with a unification scale $\sim
10^{18}{\rm GeV}$. Furthermore, decreasing the unification scale by
two orders of magnitude our bounds are increased by around 10\%.

For the one dimensional ${\bf Z}_2$ orbifold the bound for $R$ is low and
the approximation made in {\unleff} of neglecting $m_0$ is not good for
the case of $Z$-boson. Instead, using {\sleff}, we obtain:
$$
R^{-1}\simgt 150\ {\rm GeV}\ .
\equn\put\rmin
$$
On the other hand, for the two dimensional case of ${\bf Z}_3$,
${\bf Z}_4$ and ${\bf Z}_6$ orbifolds we obtain:
$$
R^{-1}\simgt 0.84\ {\rm TeV}\ ,\quad 0.68\ {\rm TeV}\ ,\quad 0.60\ {\rm
TeV} \ ,
\equn\put\rmind
$$
respectively. It is interesting that these bounds leave open the exciting
possibility of producing the lightest KK-excitations in future
supercolliders (LHC, SSC).

Finally, one of the important features of these models is that all massive
KK-modes are unstable. Here we present the computation of lifetime of the
lightest one which can decay only to two massless twisted states. It turns
out that the result remains of the same order independently of which gauge
boson of the electroweak theory one considers to have the lightest
excitation. For instance in the case of the first excitation of the photon,
$\gamma^*$, we obtain a width:
$$
\Gamma_{\gamma^*} \sim {8\over 3R} \alpha_{em}(1/R)
\sim {0.02 \over R} \ ,
\equn\put\width
$$
which leads to a lifetime of the order of $10^{-27}$ seconds for
$R^{-1}\sim 1{\rm TeV}$.

\vskip 1.0cm
\noindent{\bf Acknowledgments}
\vskip.5cm
We thank C. Bachas, J. Lascoux and particularly J. Rizos for useful
discussions. This work was supported in part by the EEC contracts
SC1-915053 and SC1-CT92-0792.

\vskip 1.5cm
\centerline{\bf REFERENCES}
\vskip 0.5cm

\parskip=-3 pt

\item{[{\cKK}]} See for example T. Appelquist, A. Chodos and P.G.O.
Freund, {\it Modern Kaluza-Klein theories},Addison-Wesley,
1987.\hfill\break

\item{[{\cGsw}]} See for example M.B. Green, J.H. Schwarz and E. Witten,
{\it Superstring theory}, Cambridge Press, 1987.\hfill\break

\item{[{\cAbl}]} T. Banks and L. Dixon, {\np} {\bf B307} (1988) 93; I.
Antoniadis, C. Bachas, D. Lewellen and T. Tomaras, {\pl} {\bf 207B}
(1988) 441; C. Kounnas and M. Porrati, {\np} {\bf B310} (1988) 355; S.
Ferrara, C. Kounnas, M. Porrati and F. Zwirner, {\np} {\bf B318} (1989)
75.\hfill\break

\item{[{\cAnt}]} I. Antoniadis, {\pl} {\bf 246B} (1990) 377; Proc.
PASCOS-91 Symposium, Boston 1991 (World Scientific, Singapore)
p.718.\hfill\break

\item{[{\cAqm}]} I. Antoniadis, C. Mu\~noz and M. Quir\'os, {\np} {\bf
B397} (1993) 515.\hfill\break

\item{[{\cKs}]} V.A. Kostelecky and S. Samuel, {\pl} {\bf270B} (1991) 21.
\hfill\break

\item{[{\cDhvw}]} L.J. Dixon, J. Harvey, C. Vafa and E. Witten, {\np}
{\bf B261} (1985) 678; {\bf B274} (1986) 285.\hfill\break

\item{[{\cGP}]} A. Giveon and M. Porrati, {\pl} {\bf 246B} (1990) 54;
{\np} {\bf B355} (1991) 422.\hfill\break

\item{[{\corb}]} S. Hamidi and C. Vafa, {\np} {\bf B279} (1987) 465;
L. Dixon, D. Friedan, E. Martinec, S. Shenker, {\np} {\bf B282} (1987)
13.\hfill\break

\item{[{\cfterm}]} M. Cvetic, {\it Phys. Rev. Lett.} {\bf 59} (1987)
1795.\hfill\break

\item{[{\cPdg}]} Particle Data Group, Review of Particles Properties,
{\it Phys. Rev.} {\bf D45} (1992) S1.\hfill\break

\item{[{\cLep}]} ALEPH Collab., D. Buskulic et al., {\it Z.
Phys.}  {\bf C59} (1993) 215.\hfill\break

\item{[{\cTas}]} TASSO Collab., W. Braunschweig et al., {\it Z.
Phys.}  {\bf C37} (1988) 171.\hfill\break

\end